\documentclass[aps,prl,preprint,groupedaddress]{revtex4}
\begin{document}
\title{ The SL(2,R)WZWN string model as a deformed oscillator and its classical-quantum string regimes}
\author{M. Ram\'on Medrano$^{1,2}$ and N. G. S\'anchez$^{2}$}
\affiliation
{(1) Departamento de F\'{\i}sica Te\'orica, Facultad de Ciencias F\'{\i}sicas, Universidad Complutense, 
E-28040 Madrid, Spain  \\
(2)Observatoire de Paris, LERMA UMR CNRS 8112, 61 Avenue de l'Observatoire, 75014 Paris, France}

\date{\today}

\bigskip

\begin{abstract}

We study the $SL(2,R)$ WZWN string model describing bosonic string theory in $AdS_3$ space-time as a deformed oscillator together with its mass spectrum
and the string modified $SL(2,R)$ uncertainty relation. The $SL(2,R)$ string oscillator is far more quantum (with higher quantum uncertainty) and more excited than the non deformed one. This is accompassed by the highly excited string mass spectrum which is drastically changed with respect to the low excited one. The highly excited quantum string regime and the low excited semiclassical regime of the $SL(2,R)$ string model are described and shown to be the quantum-classical dual of each other in the precise sense of the usual classical-quantum duality. This classical-quantum realization is not assumed nor conjectured. The quantum regime (high curvature) displays a modified Heisenberg's uncertainty relation, while the classical (low curvature) regime has the usual quantum mechanics uncertainty principle.

\bigskip

\bigskip

Norma.Sanchez@obspm.fr, mrm@fis.ucm.es 

\end{abstract}

\maketitle

\section{1. Introduction\label{sec:intro}}

 We consider the $SL(2, R)$ WZWN (Wess-Zumino-Witten-Novikov) model of level $k$, describing a bosonic string theory in a 3-dimensional Anti de Sitter space-time $(AdS_3 \cong SL(2, R) \cong SU(1, 1))$ that is, the string dynamics takes place in an exact conformally invariant background which is locally a $(2+1)$ dimensional anti de Sitter space time with torsion,\cite{1}-\cite{5}. 
 
In the algebra of the $SL(2, R)$ WZWN model, instead of having the  fundamental Abelian Harmonic oscillator commutator $\left[a, a^{+}\right] = I$, one has  a non-Abelian Kac-Moody algebra for the conserved string currents.

This algebra structure can be understood as a deformed oscillator. In particular, the $k$ dependent $SL(2, R)$ central charge leads to a precise modification of the Heisenberg's uncertainty relation in string theory, which in the limit $k \rightarrow\infty$ reduces to the usual quantum mechanical relation. The study of the WZWN model in terms of their conserved currents allows a clear characterization of string coherent states as eigenstates of the annihilation operator. These states satisfy minimal uncertainty for the generalized Heisenberg's relation, and they are in this sense the most classical ones. 

The level $k$ can be expressed as the ratio between the characteristic $AdS$ classical length $L_{c\ell}$ and the characteristic $AdS$ string length $L_s$. The inverse of these two lengths defines a classical frequency $\omega_{c\ell}$ and a string frecuency $\omega_s$ respectively.

Two different physical regimes are realized in the $SL(2, R)$ model: a strict quantum string regime and a semiclassical (quantum field theory) regime. For $k \rightarrow 0$ ($L_s\gg L_{c\ell}$), we are in  a high curvature $AdS$ quantum string regime. The algebra is, in this case, a highly deformed oscillator and the modification of the usual Heisenberg's principle is dominant in this limit. The highly excited mass spectrum is drastically modified ($m \sim n/L_{c\ell}$) with respect to the flat space one ($m^2 \sim n/\alpha'$). For $k \rightarrow \infty$ ($L_s\ll L_{c\ell}$), we are in a low curvature or semiclassical (quantum field theory) $AdS$ regime. In this limit, we have a non deformed oscillator and the usual quantum mechanics Heisenberg's principle is reproduced. Similarly, for $k \rightarrow \infty$, the flat space mass spectrum is reproduced. 

The classical and quantum $AdS$ lengths satisfy the relation $L_{c\ell}L_{s} = l_{s}^2 = \alpha'$, being $\alpha'$ the string constant and $l_{s}$ the fundamental string lentgh $(\hbar=1=c)$. The quantum string regime of the model is characterized by the quantum set ($k$, $L_s$, $\omega_s$), while the semiclassical, or quantum field theory (Q.F.T), regime is characterized by the set ($\tilde{k} = k^{-1}$, $L_{c\ell}$,  $\omega_{c\ell}$). By changing $L_{c\ell}$ into $L_s$, (and thus $k$ into $k^{-1}$), the quantum string regime becomes the semiclassical regime and conversely. Moreover, the highly excited and the low excited mass string spectra map into each other, and the same behaviors appear for the string current algebra and the Heisenberg's uncertainty relation.

 As a consequence, these two regimes are duals of each other, but in the precise sense of the usual classical-quantum (or wave-particle) duality. This classical- quantum duality does not require the existence of any isometry in the curved background nor needs a priori any symmetry nor compactified dimensions \cite{6}.

The $SL(2, R)$ WZWN model realizes, in a clear and explicit way, the dual behavior of the quantum string regime and the semiclassical (Q.F.T) regime, which is built here.

The dual nature of the relation between the quantum string regime ($QS$) and the semiclassical or quantum field theory ($QFT$) regime is also supported by the results of $QS$ and $QFT$ dynamics in curved backgrounds \cite{7}-\cite{9}.

Finally, we compare the $SL(2, R)$ string model algebra to a deformed quantum mechanics oscillator, with deformation parameter $1/k$, both models having the same modified uncertainty relations. The quantum mechanics deformed oscillator has a higher energy spectrum than the non deformed one, and analogously to the string model, for $k\rightarrow 0$ the energy excitation is maximal; on the contrary, for $k \rightarrow \infty$ (lowest excitation) the algebra, energy spectrum, and usual Heisenberg's principle of the non deformed harmonic oscillator are recovered.\\
 
\section{2. The SL(2, R) Algebra of String Currents \label{sec:alge}}
 
The SL(2, R) WZWN action has a set of conserved right and left moving currents, we call them  $J_{\pm}(\sigma^{\pm})$ ($\sigma^{\pm} = \tau\pm\sigma$ being right and left moving coordinates on the world sheet). These currents can be expressed, in terms of their Pauli components, as
 \begin{equation}
J_{\pm}= \eta_{ab}~J_{\pm}^{a}~t^{b}~=~J_{\pm}^{+}~t^{-}~+J_{\pm}^{-}~t^{+}-J_{\pm}^{3}~t^{3}
 \label{eq:jpm}
\end{equation}
being the space-time metric $\eta_{+-}=\eta_{-+}=1$,$\eta_{33}=-1$, and $t_{\pm} = \frac{1}{2}(t_1\pm it_2)$ which in terms of Pauli matrices are
\begin{equation}
t_{\pm} = \pm \frac{1}{2}(\sigma_3 \pm i\sigma_1),~~~~t_3 = \frac{1}{2}\sigma_2
\label{eq:tes}
\end{equation}

For each component of a definite current $J_{\pm}$ we have a Fourier mode expansion
\begin{equation}
J_{\pm}^{\pm,3}= \sum_{n=-\infty}^{+\infty}~ J_{\pm, n}^{\pm,3}~~e^{-i n \sigma_{\pm}}
 \label{eq:fourier}
\end{equation}

Taking, for example, just the right moving current $J_{-}$ (we drop the subindex from now on), the $SL(2,R)$ Kac-Moody algebra is
\begin{eqnarray}
\left[  J_{m}^{+}, ~J_{n}^{-}~\right] & = &\left(2~J_{m+n}^{3}~+k~ m~\delta_{m+n,0}\right) \label{eq:2Ra} \\
\left[  J_{m}^{3}, ~J_{n}^{\pm}~\right] & =&\pm~J_{m+n}^{\pm} \label{eq:2Rb}\\
\left[  J_{m}^{3}, ~J_{n}^{3}~\right] & =&-\frac{k}{2}~m~\delta_{m+n,0} \label{eq:2Rc}
\end{eqnarray}
where
\begin{equation}
k=\frac{1}{H^2~\alpha'}
 \label{eq:k}
\end{equation}
($\hbar~=~1~=~c$), $\alpha'$ is the fundamental string constant, and $ {\mathrm H} =~ \sqrt{|\Lambda|} $ where $\Lambda < 0$ is the $AdS_3$ cosmological constant. Introducing the characteristic $AdS_3$ classical length $L_{c\ell}$
\begin{equation}
L_{c\ell}= ~H^{-1}~~,
\label{eq:Lcl}
\end{equation}
and the quantum string length $L_{s}$
\begin{equation}
L_{s}=H~\alpha'~~,
\label{eq:Lq}
\end{equation}
$k$ can be expressed as the quotient of the two characteristic lengths in $AdS_3$ : 
\begin{equation}
k=\frac{L_{c\ell}}{L_{s}} 
\label{eq:kfr}
\end{equation}

We particularize now to circular string configurations on the $SL(2, R)$ group manifold . Then, the Fourier mode expansions for the components reduce to
\begin{equation}
J^{+}= J_{-1}^{+}~e^{i \sigma_{-}};~~~~J^{-}= J_{+1}^{-}~e^{-i \sigma_{-}};~~~~J^{3}= ~ J_{0}^{3}
 \label{eq:circ}
\end{equation}
(thus, the $n=0,~+1, -1$ modes are the only relevant). This subalgebra is considered for mathematical simplification and clarity reasons while describing the relevant physical contents of the model. Other string configurations can be used as well. The $SL(2,R)$ Kac-Moody subalgebra generated by the $ J_{-1}^{+}$, $J_{+1}^{-}$ and $J_{0}^{3}$ components is given by
\begin{eqnarray}
\left[  J_{+1}^{-}, ~J_{-1}^{+}~\right] & = &(~I~+ 2~\frac{J_{0}^{3}}{k}~) \label{eq:kcira} \\
\left[  J_{0}^{3}, ~J_{-1}^{+}~\right] & =&~J_{-1}^{+} \label{eq:kcirb}\\
\left[  J_{0}^{3}, ~J_{+1}^{-}~\right] & =&-~J_{+1}^{-} \label{eq:kcirc}
\end{eqnarray}
where we have used the following redefinitions

\begin{equation}
\frac{J_{+1}^{-}}{\sqrt{k}}~\rightarrow J_{+1}^{-}~~;~~\frac{J_{-1}^{+}}{\sqrt{k}}~\rightarrow J_{-1}^{+}
\label{eq:Jrdf}
\end{equation}
\\
Notice that $(~J_{-1}^{+})^{+}=~J_{+1}^{-}$ and $(~J_{0}^{3})^{+}=~J_{0}^{3}$. The $J_{+1}^{-}$ and $J_{-1}^{+}$ operator components play the role of annihilation and creation operators respectively. 

At this point, let us introduce the following self-adjoint operators 
\begin{eqnarray}
X & = &\frac{L_{s}}{\sqrt{2}} \left( J_{-1}^{+}+J_{+1}^{-}\right) \label{eq:XPa} \\
P & =& \frac{i}{\sqrt{2}~L_{s}} \left( J_{-1}^{+}-J_{+1}^{-}\right)  \label{eq:XPb}
 \end{eqnarray}
$X$ and  $P$ play the role of  position and momentum operators respectively. Then, in terms of $X$ and $P$, the algebra  Eqs.~(\ref{eq:kcira}),~(\ref{eq:kcirb}) and~(\ref{eq:kcirc}) reads:
\begin{eqnarray}
 \left[ X, P \right] &=&i~\left( I~+2~\frac{J_{0}^{3}}{k} \right)  \label{eq:xpa} \\
\left[ X, J_0^3 \right] &=&i~L_{s}^{2}~P   \label{eq:xpb} \\
\left[ P, J_0^3 \right] &=&-~i~L_{s}^{-2}~X   \label{eq:xpc} 
 \end{eqnarray}

Let us define the operator $\mathcal{H}$ in terms of the $J_{0}^{3}$ current:
\begin{equation}
\mathcal{H} = \omega_{s}J_{0}^{3} 
\label{eq:hami}
\end{equation}
where $\omega_{s}$ is the string frequency in $AdS_3$ 
\begin{equation}
\omega_{s}~= k~{\mathrm H} = L_{s}^{-1}
\label{eq:sfq}
\end{equation}

In terms of $ \mathcal{H} $,  the algebra Eqs.~(\ref{eq:kcirb}),~(\ref{eq:kcirc}), (or ~(\ref{eq:xpb}) and~(\ref{eq:xpc})) is:
\begin{eqnarray}
\left[  J_{+1}^{-}, ~J_{-1}^{+}~\right] &=&~\left( I~+\frac{2}{k~\omega_{s}}~\mathcal{H}\right)  \label{eq:JJH} \\
\left[ \mathcal{H},  J_{-1}^+ \right] &=&\omega_{s}~J_{-1}^{+}    \label{eq:HJa} \\
\left[ \mathcal{H},  J_{+1}^- \right] &=&-~\omega_{s}~J_{+1}^{-}    \label{eq:HJb} 
 \end{eqnarray}
Therefore, in terms of $X$, $P$, and $\mathcal{H}$, the algebra reads:
\begin{eqnarray}
\left[ X,  P \right] & = & i~\left(I~+\frac{2}{k~\omega_{s}}~\mathcal{H}\right)     \label{eq:XPHc} \\
\left[ X, \mathcal{H} \right] &=& i~\omega_{s}^{-1}~P     \label{eq:XPHa} \\
\left[ P, \mathcal{H} \right] & = &-i~\omega_{s}^{3}~X      \label{eq:XPHb} 
 \end{eqnarray}
 \begin{equation}
{(k\omega_{s})}^{-1}~=~\frac{L_{s}^{2}}{L_{c\ell}}= \alpha'~^{2}H^{3}
\label{eq:komegas}
\end{equation}

Notice the differences and analogies with the usual harmonic oscillator relations. Eqs.~(\ref{eq:XPHa}),~(\ref{eq:XPHb}) coincide with the usual harmonic oscillator commutation relations, while Eq.~(\ref{eq:XPHc}) is a deformed  commutator algebra due to the Kac-Moody central charge. Also, $\mathcal{H}$ here is not a quadratic Hamiltonian in the $X$ and $P$ operators, but un independent operator $ \omega_{s}J_{0}^{3}$. The deformed $SL(2,R)$ algebra leads to a precise modification of the usual Heisenberg's uncertainty relation in string theory.

\section{3. SL(2,R) quantum string uncertainty and mass spectrum \label{sec:unc}}

Let us discuss the implications for the Heisenberg's uncertainty relation in this context. We know from standard quantum mechanics that
\begin{equation}
\Delta~X_{\Psi}~\Delta~P_{\Psi} \leq \frac{1}{2}~\left| \langle \Psi|[X,P]| \Psi\rangle \right|
\end{equation}
where $|\Psi\rangle$ is a normalised quantum string state. Then, with the help of Eq.~(\ref{eq:XPHc}) we have:
\begin{equation}
\Delta X_{\Psi} \Delta~P_{\Psi} \leq \frac{1}{2}~\left(1+\frac{2}{k~\omega_s}~ \langle \Psi|\mathcal{H}| \Psi\rangle \right)
\label{eq:Heis}
\end{equation}
 
Eigenstates of the annihilation operator $ J_{+1}^{-} $, that is coherent states, satisfy minimal quantum uncertainty (equal sign in Eq.~(\ref{eq:Heis})), and are thus in this sense the most classical ones \cite{4}-\cite{5}. 

The string modified quantum uncertainty Eq.~(\ref{eq:Heis}) is \textit{larger} than the usual quantum mechanical uncertainty. The modification to the Heisenberg uncertainty relation Eq.~(\ref{eq:Heis}) is dominant in the limit $k\to0$ , that is when  $L_{s}~\gg L_{c\ell}$. In this limit,  we are strictly in the quantum string regime. On the opposite side, the classical or semiclassical regime corresponds to the limit $k \rightarrow \infty$, that is $L_{c\ell}~\gg L_{s}$, in which Eq.~(\ref{eq:Heis}) reproduces the usual quantum mechanics result. Thus,  Eq.~(\ref{eq:Heis}) shows that quantum string oscillations are far more \textit{quantum} than semiclassical (quantum field theory) oscillations $(k \rightarrow \infty)$ which in the strict limit reproduce the usual quantum mechanics uncertainty principle. 

Moreover, the study of the SL(2,R) string model in terms of their conserved currents, as a deformed oscillator, allows a clear characterization of the quantum states, namely coherent states, and the analysis of the corresponding quantum high mass spectrum\cite {4}. Quantum coherent states, eigenstates of the annihilation operator $ J_{+1}^{-} $,  generalize ordinary coherent states of quantum mechanics, and they correspond to oscillating circular strings in the classical limit \cite {5}. The quantization condition for the high mass states arises from the condition of finite positive norm of the coherent string states. In the $k \rightarrow \infty $ limit, the usual properties of coherent, or quasi-classical states, are recovered.  

Let us consider the quantum string mass spectrum of the $SL(2,R)$ model, which is given by \cite{4}
 \begin{equation}
 m^{2}\alpha'=\frac{k}{4}~+N\left( 1~+~\frac{1}{k}\right)~+~\frac{N^{2}}{k}
\label{eq:msone}
\end{equation}
There is an infinite tower of discrete mass states.
We see that, in the semiclassical o low energy regime (i.e~$k~ \rightarrow\infty$), the usual oscillators of the low excited (flat space-time)  mass spectrum are recovered:
\begin{equation}
 m^{2}\alpha'~\simeq~N~+~\frac{k}{4}
\label{eq:mstwo}
\end{equation}
This corresponds to a low curvature (low $H$) regime. The algebra (Eqs.~(\ref{eq:XPHc}),~(\ref{eq:XPHa}),~(\ref{eq:XPHb})) becomes in this case the usual (non deformed) algebra of oscillators.

On the other hand, the quantum highly excited states of the mass spectrum appear for  $~N~\gg~{k}~$ : 
\begin{equation}
 m^{2}\alpha' ~\simeq\frac{N^{2}}{k}
\label{eq:msthree}
\end{equation}
corresponding to a drastic change in the spectrum , ($ m \sim N H $), with respect to the low mass oscillators, ($ m \sim \sqrt{N/\alpha'} $). This is a high curvature (high $H$) AdS quantum regime, $L_s >> L_{cl}$. The algebra is, in this case, a highly deformed oscillator.

\section{4. Classical-quantum duality\label{sec:duality}}

The two relevant characteristic lengths $L_{c\ell}$ and $L_{s}$, Eqs.~(\ref{eq:Lcl}) and~ (\ref{eq:Lq}), satify the relation 

\begin{equation}
L_{c\ell}~L_{s}~=~l_{s}^2
\label{eq:dual}
\end{equation}

and therefore they can be related by the tilde operation
\begin{eqnarray}
 \tilde{L_{s}}& =&L_{c\ell}~=~l_{s}^{2}~L_{s}^{-1} \label{eq:duala}\\
\tilde{L_{c\ell}}&=&L_{s}~=~l_{s}^{2}~L_{c\ell}^{-1} \label{eq:dualb}
\end{eqnarray}

where $l_{s}~=~ \sqrt{\alpha'}$ is the fundamental string length, $(\hbar=1=c)$.
For the level $k$ Eq.~(\ref{eq:kfr}) and for the string frequency $\omega_{s}$ Eq.~(\ref{eq:sfq}), this imply
\begin{equation}
 \tilde{k}~=~k^{-1}
\label{eq:ktilde}
\end{equation}
\begin{equation}
\tilde{\omega_{s}}= ~ (l_{s}^{2}~\omega_{s})^{-1} = ~\omega_{c\ell}
\label{eq:dfscla}
\end{equation}
\begin{equation}
\tilde{\omega_{c\ell}}= ~ (l_{s}^{2}~\omega_{c\ell})^{-1} = ~\omega_{s}
\label{eq:dfsclb}
\end{equation}
where $\omega_{c\ell}$ is the classical frecuency in $AdS_3$
\begin{equation}
\omega_{c\ell}=L_{c\ell}^{-1}= H
\label{eq:fcl}
\end{equation}

That is, under the tilde operation, the quantum string regime becomes the semiclassical or quantum field theory regime, namely: $k \rightarrow0$ (string quantum regime) transforms into $k\rightarrow\infty$ (semiclassical regime).

This means that the quantum string limit and the semiclassical QFT limit of the $SL(2,R)$ string model are dual of each other in the precise sense of the usual classical-quantum (wave-particle) duality. In other words, the quantum string regime characterized by the set $(L_s, k, \omega_s)$ is the quantum dual of the semiclassical Q.F.T regime $(\tilde{L_s}, \tilde{k}, \tilde{\omega_s}) = (L_{c\ell}, k^{-1}, \omega_{c\ell})$. We go from one regime into the other of the same string system, both for the string current algebra, $\tilde{\omega_s}= \omega_{c\ell}$, and for the mass spectrum.

The $SL(2,R)$ string model is a well defined and explicit realization of the dual character of both regimes.

This semiclassical-quantum gravity duality does not require the existence of any isometry in the curved background, and it does not need a priori neither any symmetry nor compactified dimensions. Several types of relativistic operations $L \rightarrow \alpha'/L$ appear in string theory due to the existence of the dimensional constant $\alpha'$ (for example, $T$-duality first described on toroidal compactifications). However the duality, we are considering, is of the type classical-quantum (or wave particle) duality (de Broglie type) relating classical/semiclassical and quantum behaviors, and extended here to include the quantum string regime. The de Broglie relation $L_q = \hbar p^1$ is not the expression of a symmetry transform between physically equivalent theories, but it links, through $\hbar$, two different behaviors of nature. In a similar spirit, $L_q = l^2_{Pl}L_{c\ell}$ relates two different (semiclassical and quantum) gravity regimes.

This semiclassical-quantum realization is not \textit{assumed} nor \textit{conjectured}. Besides the explicit realization built here in the WZWN model, explicit results from the dynamics of $QFT$ and $QS$ support it in a wide class of curved backgrounds \cite{9}.

\section{Quantum mechanics deformed oscillator\label{sec:deformed}} 

Let us consider, for comparison, an oscillator hamiltonian $\mathcal{H}_{o}$ given by the usual expression
\begin{equation}
\mathcal{H}_{o}=\frac{\omega}{2}\left( a~a^{+}~+~a^{+}~a \right)
\label{eq:Hos}
\end{equation}

having the deformed commutation relations
\begin{equation}
 \left[ a,  a^{+} \right] = \left(I~+\frac{2}{k~\omega}~\mathcal{H}_{o}\right)
 \label{eq:aa}
 \end{equation}
 
Eqs.~(\ref{eq:Hos}) and~(\ref{eq:aa}) lead to: 
\begin{equation}
\left[\mathcal{H}_{o} , a\right] =-~\omega(k)\left(I~+\frac{2}{k~\omega}~\mathcal{H}_{o}\right)a 
\label{eq:ah}
\end{equation}
\begin{equation}
\left[\mathcal{H}_{o} ,a^{+}\right]=\omega(k)~a^{+}\left(I~+\frac{2}{k~\omega}~\mathcal{H}_{o}\right) \label{eq:amh} 
\end{equation}
 and $\mathcal{H}_{o}$ can be written 
\begin{equation}
\mathcal{H}_{o} = \omega(k)\left(a^{+}~a +\frac{I}{2}\right)
\label{eq:Hd}
\end{equation} 
where
\begin{equation}
\omega(k) =  \frac{\omega}{1-\frac{1}{k}}
\label{eq:wk}
\end{equation}
being $N = a^{+}a$ the number operator. Let us introduce the self-adjoint operators $X$ and $P$  
\begin{eqnarray}
X & = &\frac{L}{\sqrt{2}} \left( a^{+}+ a\right) \label{eq:Xd} \\
P & =& \frac{i}{\sqrt{2}~L} \left( a^{+}- a\right)  \label{eq:Pd}
 \end{eqnarray}

where $L = \omega^{-1}$ is the oscillator length, then Eqs.~(\ref{eq:Xd}),~(\ref{eq:Pd}) and~(\ref{eq:aa}) lead to the following algebra
\begin{eqnarray}
 \left[ X, P \right] &=& i~\left(I~+\frac{2}{k~\omega}~\mathcal{H}_{o}\right) \label{eq:con}\\
\left[ X, \mathcal{H}_{o} \right] &=&i~L^2~\left(\omega~P + k^{-1}\left\{\mathcal{H}_{o} , P \right\} \right)   \label{eq:conx} \\
\left[ P, \mathcal{H}_{o} \right] &=&-~\frac{i}{L^2}~\left(\omega~X +  k^{-1} \left\{ \mathcal{H}_{o}, X \right\} \right)  \label{eq:conp} 
 \end{eqnarray}
 
from which the conmutators $ \left[ X, \mathcal{H}_{o} \right]$ and $\left[ P, \mathcal{H}_{o} \right]$ can be solved and the full algebra can be written as:
\begin{eqnarray}
 \left[ X, P \right] &=& i~\left(I~+\frac{2}{k~\omega}~\mathcal{H}_{o}\right) \label{eq:cona}\\
\left[ X, \mathcal{H}_{o} \right] &=&~~\frac{-i~\omega^{-1}}{(1-k^{-2})}~\left(I~+\frac{2}{k~\omega}~\mathcal{H}_{o}\right)  ~\left( \frac{\omega ^{-2}X}{k}  + i~P\right)\label{eq:conxb} \\
\left[ P, \mathcal{H}_{o} \right] &=&~~\frac{-i~\omega^{3}}{(1-k^{-2})}~\left(I~+\frac{2}{k~\omega}~\mathcal{H}_{o}\right)  ~\left( X  + i~ \frac{\omega^{-2} P}{k}\right)  \label{eq:conpc} 
 \end{eqnarray}

For $ k\rightarrow\infty$ the algebra reduces to the usual harmonic oscillator algebra. 

This algebra (Eqs.~(\ref{eq:cona})-(\ref{eq:conpc})) can be compared to the $SL(2,R)$ string oscillator algebra described above. The $\left[ X, P \right]$ commutators  of both systems are the same, and the two other commutators coincide in the $ k\rightarrow\infty$ limit.  
 
 From Eq.~(\ref{eq:Hd}), the energy spectrum is given by
 \begin{equation}
 E_{n}~=~\omega(k) \left( n~+~\frac{1}{2}\right) 
 \label{eq:Ek}
 \end{equation}
Since $\omega(k) > \omega$, the deformed oscillator has a {\it higher} energy spectrum than the non deformed one. In particular, the constant energy gap is larger than the one in the normal oscillator. The energy excitation is maximal in the $k \rightarrow\ 0$ limit. On the contrary, for $k\rightarrow\infty$, $\omega(k)\rightarrow \omega$, the algebra and the energy spectrum become the ones for the non deformed oscillator. The usual non deformed oscillator is a low excited regime with respect to the deformed one. 

Thus, the SL(2,R) string model can be compared to a deformed oscillator of string frequency $\omega_s$ and deformation parameter $1/k$.  The $SL(2,R)$ string algebra Eqs.~(\ref{eq:JJH})- (\ref{eq:HJb}), where $J_{+1}^{-}$ and $J_{-1}~^{+}$ are annihilation and creation operators respectively,  can be compared to the  Eqs.~(\ref{eq:aa})-~(\ref{eq:amh}) for the deformed oscillator. Both models have the same modified uncertainty relations, although for the $SL(2,R)$ string model the hamiltonian  $\mathcal{H}$ is not a function of  the $J_{+1}^{-}$ and $J_{-1}~^{+}$ operators but an independent operator $J_{0}^{3}$ (Eq.~(\ref{eq:hami})). The energy spectrum of the deformed oscillator is far more excited than the non deformed one, as the mass spectrum of the SL(2,R) string deformed oscillator is far more quantum than the low excited $k\rightarrow\infty$ (semiclassical) one.

\bigskip

\begin{acknowledgements}

{\bf Acknowledgements}

\bigskip
 
M. R. M. acknowledges the Spanish Ministry of Education and Science (FPA04- 2602) for financial
support, and the Observatoire de Paris, LERMA, for the kind hospitality extended to her.
\end{acknowledgements}


\begin{thebibliography}{}

\bibitem{1} J. Balog, L. O' Raifeartaigh, P. Forgacs and A. Wipf, \emph{Nucl. Phys.{ \bf B325}; 225} (1989); E. Witten, \emph{Commun. Math. Phys. {\bf 92}, 455} (1984);
L.D. Dixon, M.E.Peskin and J. Lykken, \emph{Nucl. Phys. {\bf B325}, 329} (1989);
N. Mohammedi, \emph{Int. Journ. Mode. Phys. {\bf A5}, 3201} (1990);
I. Bars and D. Nemeschansky, \emph{Nucl. Phys. {\bf B348},89 } (1991); S. Hwang, \emph{Nucl. Phys. {\bf B354}, 100} (1991); \emph{Phys. Lett.{\bf B334}, 451} (1992).

\bibitem{2} H. J. de Vega, A. L. Larsen and N. S\'anchez, \emph{Phys.Rev. {\bf D58}, 26001} (1998).

\bibitem{3}A. L. Larsen and N. S\'anchez, \emph{Phys.Rev. {\bf D58}, 126002} (1998). 

\bibitem{4} A. L. Larsen and N. S\'anchez, \emph{Phys.Rev. {\bf D62}, 46003} (2000). 

\bibitem{5} A. L. Larsen and N. S\'anchez, \emph{Nucl. Phys. {\bf B618}, 301} (2001). 

\bibitem{6}  
  E. \'Alvarez, L. \'Alvarez-Gaum\'e and Y. Lozano, \emph{Phys. Lett. {\bf B336}, 183} (1994); \emph{Nucl.Phys.Proc.Suppl.{ \bf41}, 1} (1995).

\bibitem{7} M. Ram\'on Medrano and N. S\'anchez, \emph{Mod. Phys. Lett. {\bf A18}, 2537} (2003).

 \bibitem{8} N. S\'anchez, \emph{Int. J. Mod. Phys. {\bf A19}, 4173} (2004). 
 
  \bibitem{9} M. Ram\'on Medrano and N. S\'anchez, \emph{Phys.Rev. {\bf D60}, 125014} (1999); M. Ram\'on Medrano and N. S\'anchez, \emph{Phys.Rev. {\bf D61}, 084030}(2000); 
A. Bouchareb, M. Ram\'on Medrano and N. S\'anchez, hep-th/0508178, hep-th/0511281. 

\end{thebibliography}
\end{document}